\documentclass[traditabstract]{aa}
\usepackage{epsfig,graphicx}
\usepackage{natbib}
\bibpunct{(}{)}{;}{a}{}{,}    % to follow the A&A style

\begin{document}

\title{FTS atlas of the Sun's spectrally resolved
  center-to-limb variation}
%\subtitle{}
\titlerunning{FTS atlas of the Sun's spectrally resolved
  center-to-limb variation}
\author{J.O. Stenflo\inst{1,2}}% \and
\institute{Institute of Astronomy, ETH Zurich, CH-8093 Zurich \and
  Istituto Ricerche Solari Locarno, Via Patocchi, CH-6605 Locarno
  Monti, Switzerland} 

\date{}

%%%%%%%%%%%%%%%%%%%%%%%%%%%%%%%%%%%%%%%%%%%%%%%%
\abstract{The Sun's spectrum varies with center-to-limb distance,
  which is usually parameterized by $\mu=\cos\theta$, where $\theta$ is
  the heliocentric angle. This variation is governed by the underlying
  temperature-density structure of the solar atmosphere. While the
  center-to-limb variation (CLV) of the continuous spectrum is well known
  and has been widely used for atmospheric modeling, there has been no
  systematic exploration of the spectrally resolved CLV. Here we make
  use of two spectral atlases recorded with the Fourier transform
  spectrometer (FTS) at the McMath-Pierce facility at Kitt Peak. One
  spectral atlas obtained 10\,arcsec inside the solar limb was recorded in
  1978-79 as part of the first survey of the Second Solar
  Spectrum, while the other atlas is the well used reference NSO/Kitt Peak FTS atlas for
  the disk center. Both atlases represent fully resolved spectra without any
  spectral stray light. We then construct an atlas of the limb/disk-center
  ratio between the two spectra over the wavelength range
  4084-9950\,\AA. This ratio spectrum, which expresses the CLV amplitude relative to
  the continuum, is as richly structured as the intensity
  spectrum itself, but the line profiles differ greatly in both shape and
  amplitude. It is as if we are dealing with a new, unfamiliar
  spectrum of the Sun, distinctly different from both the intensity
  spectrum (which we here refer to with the acronym SS1) and the
  linear polarization of the Second Solar Spectrum (for which we use 
  acronym SS2). In analogy we refer to the new ratio
  spectrum as SS3. While there is
  hardly any resemblance between SS3 and SS2, we are able to identify
  a non-linear mapping that can translate SS1 to SS3 in the case of
  weak to medium-strong spectral lines that are mainly formed in LTE
  (being directly coupled to the local temperature-density
  structure). This non-linear mapping is successfully modeled in terms
  of two free parameters that are found to vary approximately linearly over
  the entire wavelength range covered. These parameters and the
  various SS3 line profiles provide a novel, rich set of observational
  constraints, which may be used to test the validity of model
  atmospheres or guide the construction of improved models. 
%context
%{Here comes the abstract.}
%aims
%{} 
%methods
%{}
%results
%{}
%conclusions
%{}

}

\maketitle

%%%%%%%%%%%%%%%%%%%%%%%%%%%%%%%%%%%%%%%%%%%%%%%%%%%%%%%%%%%%%
\section{Introduction}\label{sec:intro}
The main reason why the Sun's disk is limb darkened at visible
wavelengths is because the temperature decreases with height in the
layers where the spectrum is formed. The center-to-limb position is
usually characterized by $\mu$, the cosine of the heliocentric angle
$\theta$. The radiation that reaches us from different $\mu$ positions
comes from different atmospheric heights. On average the photons that
we receive emanate from an optical depth $\tau =1$ along the line of
sight, which corresponds to an optical depth in the vertical direction
$\tau_v =\mu$. Because the conversion of the optical depth scale
$\tau_v$ to a geometrical height scale $h$ depends on the opacity,
which is a strong function of wavelength, the limb darkening also
varies with wavelength and becomes increasingly pronounced as we go
down in wavelength into the UV range. The coupling between limb
darkening and temperature gradient is not complete due to non-LTE
effects related to scattering processes, which have no direct relation
to temperature, but these effects play a secondary role for continuum
radiation. 

Due to the close relation between limb darkening and the temperature
and opacity structure, the observed limb darkening with its wavelength
dependence constitutes one of the main constraints on modeling of the
Sun's atmosphere. However, so far it has mainly been the continuum limb
darkening that has been used, e.g. as determined by
\citet{stenflo-neckellabs94}, and not the limb-darkening profiles of
the various spectral lines \citep[cf.][]{stenflo-asplund09}. 

The limb darkening however has a rich spectral structure, since the
opacity, height of formation, and non-LTE effects vary across the
profiles of each of the numerous spectral lines. The spectrally
resolved limb darkening gives us an enormously richer set of
observational constraints on the structure of the solar atmosphere as
compared with the continuum limb darkening alone. The aim of the present
paper is to fully spectrally resolve the way in which the Sun's spectrum varies
between disk center and the limb, across the spectral range 4084 --
9950\,\AA, and to discuss the properties and meaning of this
new kind of spectrum. 

The limb darkening can also be interpreted in terms of the angular
distribution of the emergent radiation at the Sun's surface. The
variation with $\mu$ across the disk also represents the angular
variation of the intensity with angle $\theta$ relative to the
vertical direction ($\mu=\cos\theta$). There would be no limb
darkening if the radiation field were isotropic within the outwards
half sphere. The presence of limb darkening thus implies a radiation field
that is more intense in the vertical than in the inclined directions. 

It is this anisotropy that is the source of the scattering
polarization that is referred to as the Second Solar Spectrum \citep{stenflo-sk97}. The
anisotropy breaks the symmetry that enables the scattered light to
become polarized. The Second Solar Spectrum, which we 
will here refer to with the acronym SS2, is as spectrally structured as the
ordinary intensity spectrum (for which we use the acronym SS1), but
the appearance of the spectrum is totally different, 
because the spectral structures are governed by different physical
processes. We will find that the center-to-limb variation
(CLV) is similarly richly structured, but in ways that differ
profoundly from both SS2 and SS1. It is therefore like having
uncovered another, previously unfamiliar spectral face of the Sun, 
which we like to think of as the Third
Solar Spectrum, but here we more conveniently choose to refer to it with the acronym
SS3. In the present paper we will compare 
SS3 with SS1 and SS2 and discuss to what extent they are physically
related.

%%%%%%%%%%%%%%%%%%%%%%%%%%%%%%%%%%%%%%%%%%%%%%%%%%%%%%%%%%%%%
\section{FTS atlas of the limb spectrum}\label{sec:ftslimb}
In the 1970s a remarkable and uniquely powerful instrument was
developed for use at the McMath-Pierce facility at Kitt Peak, the
Fourier transform spectrometer FTS
\citep[cf.][]{stenflo-brault78,stenflo-brault85}. It allows the Sun's 
fully resolved spectrum to be recorded simultaneously for all the
wavelengths within the range of the broad prefilters used (typical
band width 1000\,\AA\ or more) with high S/N ratio. There is no
significant spectral 
broadening, no spectral stray light, and the continuum level is very
well defined due to the simultaneous extremely broad spectral coverage. 

The spectral atlas recorded with this FTS at disk center
\citep{stenflo-ftsatlas84} is well known 
and has been abundantly used as the standard reference atlas. A digital
version compiled by H. Neckel, Hamburg, is publicly available. It
covers the range 3290-12508\,\AA\ in terms of an equidistant wavelength scale
with 5\,m\AA\ increments. 

It is less well known that a corresponding limb atlas exists, which
was recorded with the same FTS in the context of making the first survey of the
Second Solar Spectrum SS2 \citep{stenflo-setal83b}. In polarimetric mode,
with electro-optic modulation, lock-in amplifier, and a heterodyning
technique, the spectra of two Stokes parameters (the intensity and one
of $Q$, $U$, or $V$) could be recorded simultaneously, which allowed
atlases of the polarized spectrum to be made. While the $Q/I$ linear
polarization spectrum recorded at that time is now only of historical
interest, since modern polarimeters like ZIMPOL have a polarimetric
precision ($10^{-5}$) that is about two orders of magnitude better,
the intensity spectrum recorded simultaneously with the FTS polarimeter
is of similar superb quality as the reference disk center atlas,
namely fully spectrally resolved, no spectral stray light, high S/N
ratio, and well defined continuum level. 

This limb FTS atlas \citep[described in][]{stenflo-setal83b} was recorded on
October 2-3, 1978, and April 27-28, 1979, and covers the range
4084-9950\,\AA. The spatial field of view 
was a rectangular 17.5\,arcsec $\times$ 10\,arcsec aperture centered
10\,arcsec inside the solar limb near one of the heliographic poles
(to minimize the effect of magnetic fields). The 17.5\,arcsec side was
parallel to the limb, so the limb distance interval covered is
5-15\,arcsec. At 10\,arcsec limb distance, $\mu =0.144$. The
effective, average $\mu$ depends on the intensity weighting through
the limb-darkening function. If we for instance use 
the limb darkening that is valid for the continuum at 5000\,\AA, then
we find an intensity-weighted average $\mu$ of 0.145. This example 
demonstrates that this average is insensitive to the choice of limb
darkening function. The $\mu$ average is further insensitive to
seeing, because seeing mainly broadens the effective aperture but
does not significantly shift it. Of greater concern is the exact
positioning of the aperture by the guiding system and the stability
of the guiding (with respect to slow drifts) during the typically
1-hr long integration time used to record the FTS interferogram with each
1000\,\AA\ prefilter. Although it is hardly possible to give a good
value for the 1-$\sigma$ error in the average $\mu$ position, we 
estimate that the error is at least 0.01 in $\mu$. While a third
decimal in $\mu$ is therefore not significant, we choose to retain it here to
keep our best estimate centered, and adopt 
$\mu = 0.145$ as representing the nominal disk position to which the
limb spectrum refers. 

Since the reference disk spectrum and the limb spectrum were recorded
at different times, they can be affected by small differential Doppler
shifts due to different velocities between the telescope and the Sun
(in particular due to the rotation and orbital elliptical motion of
the Earth). Therefore we have used the disk center FTS spectrum as a
reference and made the wavelength scale of the limb spectrum conform
to the scale of this  disk center spectrum by iterative least squares
fitting of the two spectra, using shift and stretch as the two free
parameters of the fit. Only when the two spectra refer to the
identical wavelength scale is it possible to form ratios between them,
as we will do in the next section. Errors in the matching of the two
wavelength scales will produce spurious 
antisymmetric profile shapes in the ratio spectrum. The absence of
such spurious features serves as verification that the matching of the
two spectra has been successful. 

Note however that when fitting the wavelength
  scale of the limb spectrum to the scale of the disk center spectrum,
  one loses one of the characteristics of these spectra. Because of
  small-scale inhomogeneities of the solar atmosphere, in particular
  brightness-velocity correlations in the solar granulation, the
  spectral lines become asymmetric and slightly shifted in a way that
  varies with center-to-limb distance and depends on the temperature
  sensitivity and depth of formation of each spectral line. These
  subtle individual line shifts cannot be explored with the 
  present data set, which instead is used here to study the CLV of the
  depths and widths of the spectral lines, properties that are
  conveniently expressed in terms of the ratio spectrum SS3.

%%%%%%%%%%%%%%%%%%%%%%%%%%%%%%%%%%%%%%%%%%%%%%%%%%%%%%%%%%%%%
\section{Atlas of the center-to-limb variations}\label{sec:clvatlas}

\subsection{Defining CLV relations}\label{sec:define}
Let us define the center-to-limb variation function as 
\begin{equation}\label{eq:clvdef}
{\cal C}_\lambda(\mu)=I_\lambda(\mu)/I_\lambda(1.0)\,,
\end{equation}
the intensity at the given wavelength $\lambda$ as a function of
$\mu$, normalized to the corresponding intensity at disk center
($\mu=1.0$). The CLV function that represents the
continuum at this wavelength will be denoted ${\cal C}_{c,\, \lambda}(\mu)$. 

The observed spectrum is however not represented by $I _\lambda$,
which is the intensity in absolute units, but
by the rest intensity 
\begin{equation}\label{eq:restint}
r_\lambda=I_\lambda/I_{c,\, \lambda}\,,
\end{equation}
the intensity normalized to the intensity of the continuum. Let us
define the ratio between the so normalized spectra, recorded at
positions $\mu$ and disk center, respectively, as 
\begin{equation}\label{eq:capr}
R_\lambda(\mu)=r_\lambda(\mu)/r_\lambda(1.0)\,.
\end{equation}
By definition the continuum value for $R$ is always unity. Then the CLV of Eq.~(\ref{eq:clvdef}) becomes 
\begin{equation}\label{eq:crc}
{\cal C}=R\,\,\,{\cal C}_c\,,
\end{equation}
where all the three factors depend on $\lambda$ and $\mu$ (but here not
explicitly shown, for notational simplicity). Eq.~(\ref{eq:crc}) shows that the
ratio spectrum $R_\lambda(\mu)$ needs to be scaled
with the CLV of the continuum to give us the full
CLV. $R_\lambda(\mu)$ by itself represents the CLV at wavelength 
$\lambda$ {\it relative to the CLV of the continuum}. The spectral
structures seen in $R_\lambda(\mu)$ represent the {\it differential effects}
of the CLV between the lines and the continuum. 

%%%%%%%%%%%%%%%%%%%%%%%%%
\subsection{Limitations of the present atlas and its extension at IRSOL}\label{sec:atlaslim}
In the present paper we present the spectrally fully resolved FTS
atlas  of the ratio spectrum $R_\lambda$ over the range 4084-9950\,\AA\ for the
single center-to-limb position $\mu=0.145$. It represents the ratio between the
continuum-normalized FTS spectra at this limb position (10\,arcsec
inside the limb) and at disk center. We do not have FTS spectra for other
$\mu$ positions that would allow us to reproduce the {\it shape} of
the $\mu$ variation of $R$ for each wavelength 
from unity at disk center to the value observed at limb position
$\mu=0.145$. Our ratio spectrum $R_\lambda(\mu=0.145)$ however gives
us the relative {\it amplitude} of the CLV variation. 

To find the $\mu$ dependence as a function of 
wavelength requires a new observational atlas project. Such a project
is currently being carried out and is close to completion at IRSOL
(Locarno, Switzerland), with the help of recently implemented computer control of
the spectrograph and telescope to allow automatic recording of the
various spectral sections for a sequence of $\mu$ positions
(M. Setzer, M. Bianda, R. Ramelli, private communication). This
forthcoming work will complement the present FTS atlas to provide the
missing information on the relative $\mu$ variation. The shape of the
$\mu$ function is needed when one for instance wants to integrate over
the $\mu$ dependence to obtain the radiation anisotropy factor, which 
governs the symmetry breaking that is at the origin of the Second
Solar Spectrum (cf. Sect.~\ref{sec:rel2SS2} below).

\begin{figure}
\resizebox{\hsize}{!}{\includegraphics{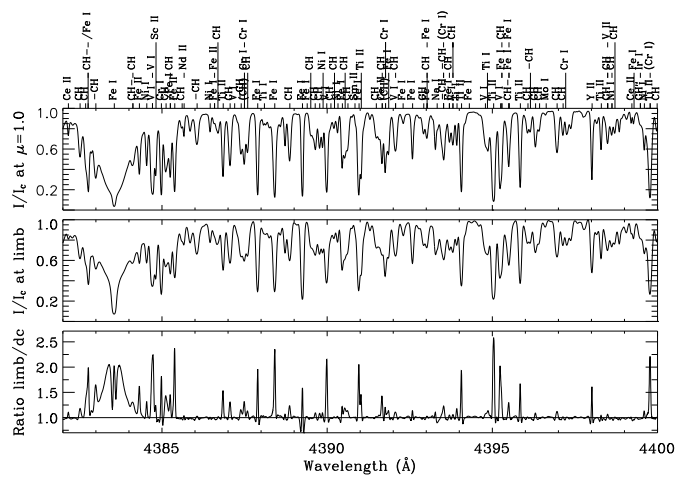}}
\caption{Example from the FTS atlas of the CLV. The top panel shows
  the disk-center spectrum with line identifications, the middle panel
  the limb spectrum (at 
$\mu=0.145$), both normalized to the level of the continuum. The 
bottom panel gives the ratio $R$ between the limb and disk-center
spectra. By definition the continuum is represented by the level
unity. Amplitudes above unity imply that the limb darkening inside the
respective lines is {\it flatter} than in the continuum. 
}\label{fig:4391}
\end{figure}

The atlases of the limb spectrum, disk-center spectrum, and ratio
spectrum $R$ are available both as pdf and as data files at the IRSOL
web site www.irsol.ch. Figure \ref{fig:4391} illustrates the rather
typical appearance of the structuring of these three spectra, here
shown for the range 4382-4400\,\AA. The top panel is from the
disk-center atlas, the middle panel for the limb atlas (at
$\mu=0.145$), while the bottom panel represents the ratio $R_\lambda$
between the limb and disk-center spectra. 

%%%%%%%%%%%%%%%%%%%%%%%%%
\subsection{Properties of the limb/disk-center ratio spectrum}\label{sec:proprat}
The first striking property of the ratio spectrum $R$ as revealed by
Fig.~\ref{fig:4391} is that nearly all the spectral structures shoot
up above the level unity that represents the continuum, there is not
much that dips below this level. Since the total limb darkening ${\cal C}$ according to
Eq.~(\ref{eq:crc}) is obtained through multiplication of $R$ with the
continuum CLV ${\cal C}_c$, which decreases rather steeply 
towards the limb, the high values of $R$ imply that
the continuum limb darkening gets raised to make the total limb
darkening become much {\it flatter} inside the spectral lines. 
Flatter limb darkening implies smaller radiation-field
anisotropy, i.e., diminished source of scattering
polarization. 

The second striking property is that the weaker lines in the intensity
spectrum are greatly suppressed or absent in the $R$ spectrum, and
that the $R$ spectrum peaks are systematically much narrower than the
corresponding intensity profiles, with the exception of strong lines
with pronounced damping wings, like the Fe\,{\sc i} line at
4383.5\,\AA. Apart from such strong lines, the moderately stronger
lines (without well developed damping
wings) get amplified in the $R$ spectrum while the
weaker lines get suppressed, which implies a relation
between $R$ and the intensity spectrum that is highly non-linear. In
Sect.~\ref{sec:model} below we will model this non-linear relation. 

\begin{figure}
\resizebox{\hsize}{!}{\includegraphics{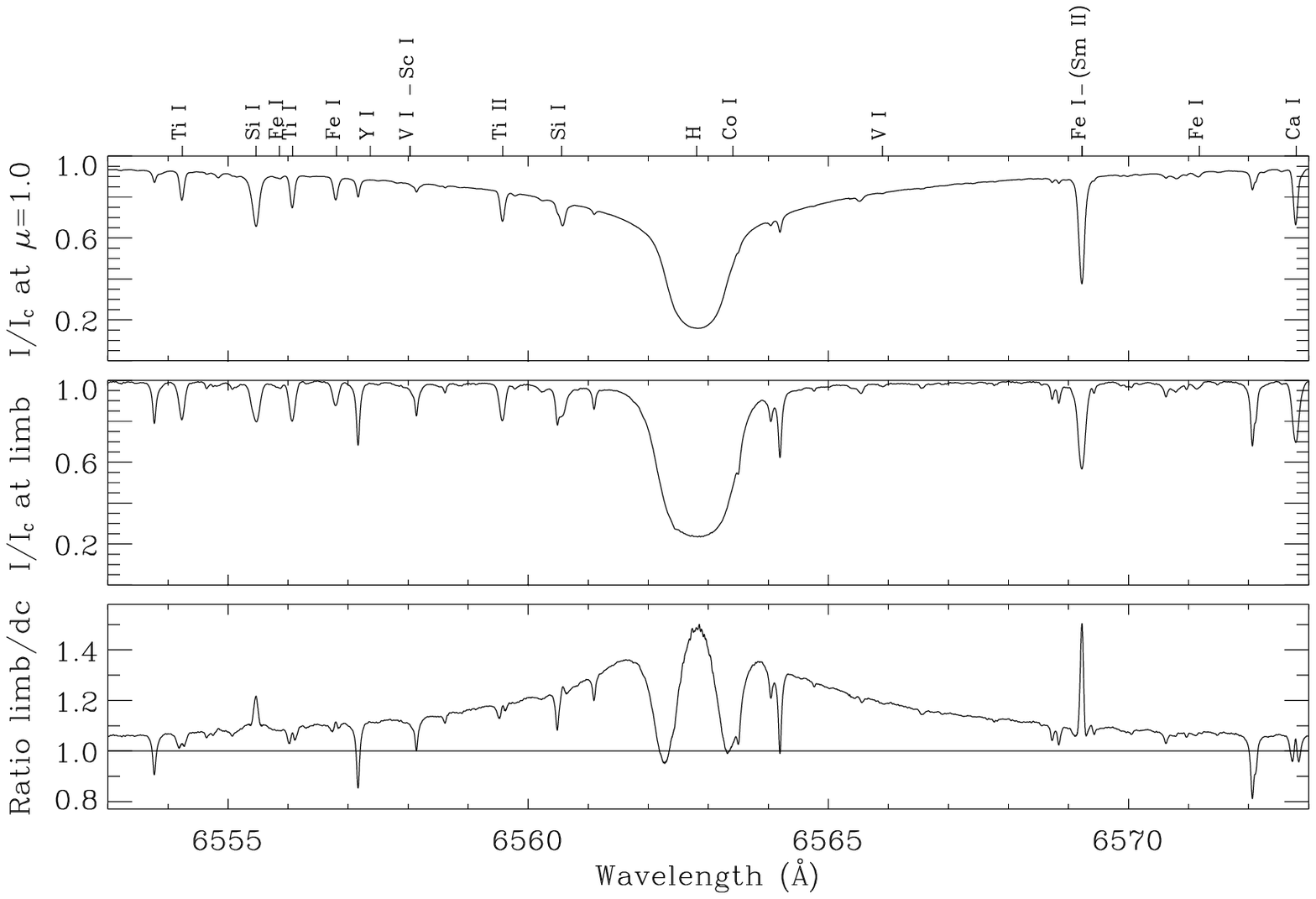}}
\caption{Section of the FTS atlas around the hydrogen H$\alpha$
  6563\,\AA\ line. Like in Fig.~\ref{fig:4391} the panels represent
  (from top to bottom) the disk-center spectrum, the limb spectrum,
  and the ratio $R$ between the limb and disk-center spectra. Note in
  particular the differential behavior between disk center and 
  limb of the extended damping wings. 
}\label{fig:6563}
\end{figure}

\begin{figure}
\resizebox{\hsize}{!}{\includegraphics{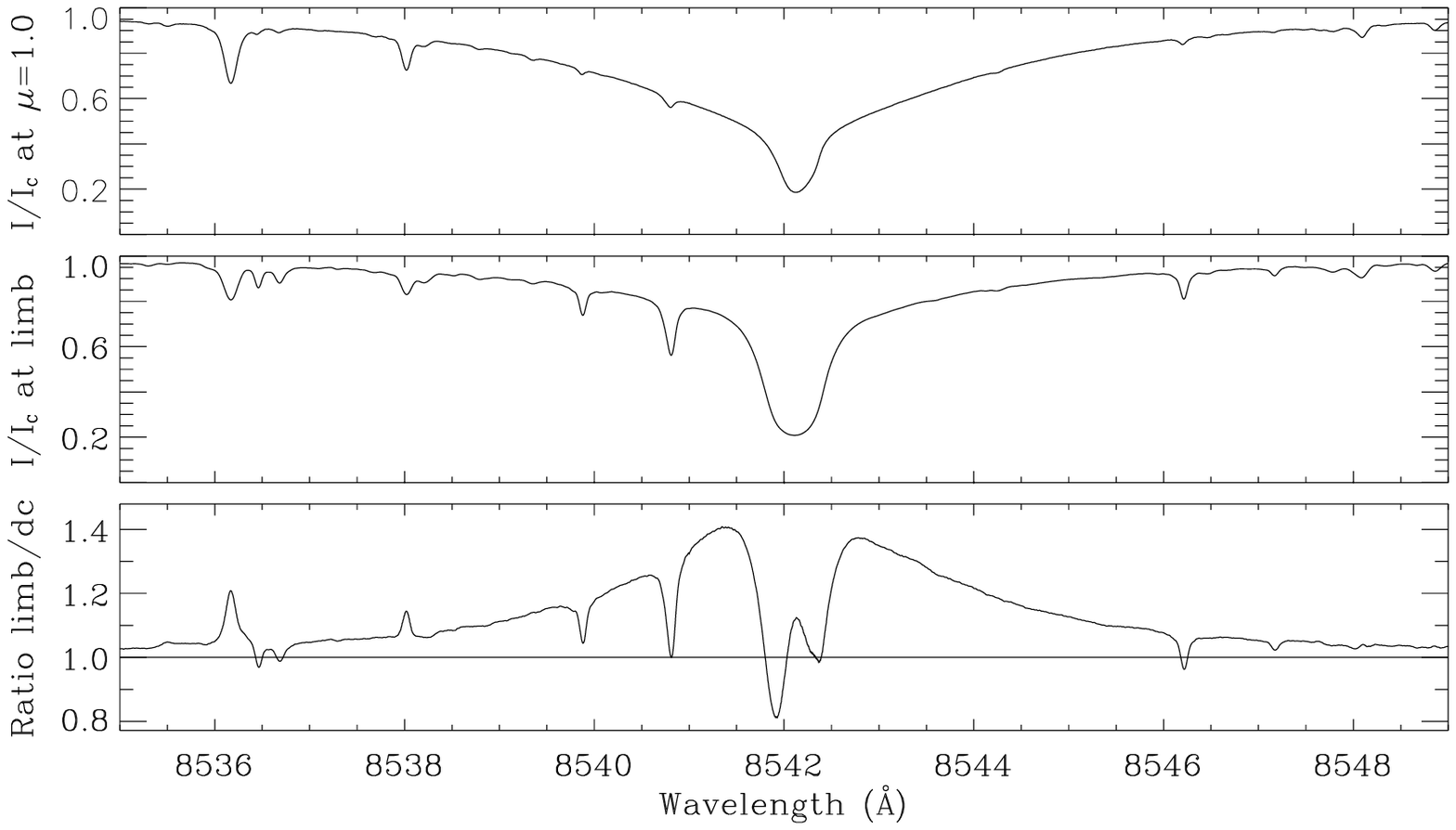}}
\caption{Section of the FTS atlas around the Ca\,{\sc
  ii} infrared triplet line at 8542\,\AA. The panels are the same as
in Figs.~\ref{fig:4391} and \ref{fig:6563}. 
}\label{fig:8542}
\end{figure}

The kind of lines for which such non-linear modeling seems to work are
the ones that can be described well in terms of LTE, lines which are closely
coupled to the temperature structure of the atmosphere. As the values of
the parameters of the non-linear model in Sect.~\ref{sec:model} 
directly depend on the temperature-density structure of the
atmosphere, they may serve as convenient new constraints on 
model atmospheres. The behavior is quite different for strong lines with
damping wings, in particular those whose line cores are 
formed in the chromosphere. As two examples we show in
Figs.~\ref{fig:6563} and \ref{fig:8542} two particularly well-known
lines: the Balmer hydrogen H$\alpha$ 6563\,\AA\ line, and the Ca\,{\sc
  ii} infrared triplet line at 8542\,\AA. 

It is particularly striking in the case of the H$\alpha$ line that the
extended damping wings that are so prominent in the disk-center
spectrum are almost absent in the limb spectrum, while the central
Gaussian-like core is greatly widened in the limb spectrum. This has
the consequence that the ratio spectrum that maps the differential
effects has a core peak and very extended wings. While the
quantitative aspects of this behavior depend on the
temperature-density structure of the Sun's atmosphere, the profile
shapes are also governed by non-LTE physics with partial frequency
redistribution (PRD). 

The behavior of the 8542\,\AA\ line is qualitatively similar, although the
suppression of the damping wings is less complete in the limb
spectrum, while the core peak in $R$ is small and asymmetric (possibly
due to a miniscule blend line). As a consequence of the non-LTE effects
all these strong lines have their own individual CLV
behavior. Nevertheless one should ideally be able to
quantitatively reproduce the CLV profile behavior of {\it 
  all} the various lines in the solar spectrum with a  {\it single} model
atmosphere (which might be multi-dimensional, which would then require horizontal
spatial averaging of the emergent radiation before
comparison with the observational constraints). All the various $R$ 
profiles in our spectral atlas collectively impose an enormously rich
set of constraints that should in principle be satisfied if the 
model is to represent the spatially averaged quiet Sun.

%%%%%%%%%%%%%%%%%%%%%%%%%%%%%%%%%%%%%%%%%%%%%%%%%%%%%%%%%%%%%
\section{Relation to the Second Solar Spectrum}\label{sec:rel2SS2}

\subsection{Role of the radiation-field anisotropy}\label{sec:clvanis}
To obtain a general, conceptual understanding of the role of the limb
darkening function for the linear polarization $p$ that is
generated by coherent scattering processes, we formally express
$p$ in the factorized form \citep{stenflo-s82} 
\begin{equation}\label{eq:pfact}
p=\alpha\,W_{2,\,{\rm eff}}\,k_G\,k_c\,k_H\,,
\end{equation}
where $\alpha$ is the fraction of the received photons that are part
of coherent scattering processes, and $W_{2,\,{\rm eff}}$ is the effective,
intrinsic atomic polarizability governed by the quantum mechanics of
the scattering system. While the intrinsic polarizability $W_2$ is a constant for
each given atomic transition, determined by the quantum numbers of the
atomic levels that are involved in the scattering process, we have
added {\it eff} in the index position to formally account for the
numerous cases when there are quantum interferences between widely
separated scattering transitions, which causes $W_2$ to be strongly
wavelength dependent across the line profiles. $k_c$ is a collisional
depolarization factor, which is unity in the absence of
collisions. Similarly, $k_H$ is the Hanle depolarization factor,
which is unity in the absence of magnetic fields. 

The remaining factor, $k_G$, is a geometric depolarization factor,
defined as follows: For single scattering at $90^\circ$ by a classical
dipole oscillator (for which $W_2=1$) in the absence of collisions and
magnetic fields, $p$ would be 100\,\%\ (the scattered radiation being fully
linearly polarized with the electric vector perpendicular to the plane
of scattering). On the Sun this case would occur for a classically
scattering system at the extreme solar limb ($\mu=0$) if all the
incident radiation would come in the vertical direction. Such a situation
represents extreme limb darkening, when the disk
center is much brighter than anything else. The electric vector of the
scattered radiation will then be oriented parallel to the limb. This
singular case with directed radiation corresponds to $k_G=1$. 

In reality the degree of limb darkening is modest and varies with
wavelength. Without limb darkening (flat disk) $k_G$ would be zero. As
$k_G$ represents the geometric symmetry breaking that is the source of
the polarization, $k_G=0$ at disk center due to axial symmetry
there. It varies with disk position (defined by $\mu$) as follows
\citep{stenflo-s82}
\begin{equation}\label{eq:kg}
k_G=G\,(1-\mu^2)\,/I_\lambda(\mu)\,,
\end{equation}
where $I_\lambda$ is the intensity of the solar disk at position
$\mu$. The factor $G$ is a measure of the anisotropy of the radiation
field: 
\begin{equation}\label{eq:g}
G={3\over 8}\,\int\,{{\rm d}\Omega\over 4\pi}\,(3\cos^2\theta
-1)\,I_\lambda(\theta,\,\varphi)\,,
\end{equation}
where $\theta$ and $\varphi$ are the colatitude and azimuth angles of the
incident radiation field. 

In the formalism of irreducible spherical tensors \citep{stenflo-lanlan04} the anisotropy of the radiation field is expressed in terms of
the tensor $J_0^2(\nu)$. It is proportional to $G$, and the
proportionality factor is close to unity: 
\begin{equation}\label{eq:j02}
G\,/\,J_0^2(\nu)\,={3\sqrt{2}\over 4}\approx 1.06\,.
\end{equation}

In Eqs.~(\ref{eq:kg}) and (\ref{eq:g}) the units for the intensity
have not been specified, but whatever units are used, they divide out
when forming $k_G$, which is dimensionless. If we normalize the
intensity to its value at disk center, then $I_\lambda$ becomes
dimensionless and equal to the center-to-limb function ${\cal
  C}_\lambda(\mu)$ defined by Eq.~(\ref{eq:clvdef}). Then also $G$
becomes dimensionless. If the radiation field is axially symmetric
around the vertical direction, which is the case for a limb-darkened,
spherically symmetric Sun, we can integrate away the azimuth angle
$\varphi$ to obtain 
\begin{equation}\label{eq:gaxisym}
G={3\over 16}\,\int\,(3\mu^{\prime 2} -1)\,{\cal
  C}_\lambda(\mu^\prime)\,{\rm d}\mu^\prime\,.
\end{equation}
For clarity we here mark the incident directions with a prime
($\mu^\prime$), to distinguish from the symbol for the disk position
($\mu$ in Eq.~(\ref{eq:kg})) from where the observed photons
come. Note that we only integrate over the outwards half sphere, since ${\cal
  C}_\lambda(\mu^\prime)=0$ for negative $\mu^\prime$ at the surface
of the Sun. 

Using the CLV for the continuum as tabulated by \citet{stenflo-pierce00}, $k_G$
has been computed and plotted as a function of wavelength from 3000 to
7000\,\AA\ in \citet{stenflo-s05}. 

In the present paper we cannot compute $G$ and $k_G$, which would
require integration over all $\mu$ from 0 to 1, because we only have
spectral atlases for two $\mu$ positions (1.0 and 0.145). Such
computations are planned for a follow-up project that will become possible
after the currently ongoing CLV atlas project with ZIMPOL at IRSOL has
been completed. The IRSOL project will cover the whole $\mu$ range
with 10 $\mu$ positions from $\mu=0.1$ to 1.0 in increments of
0.1. However, our present 
limb over disk-center ratio spectrum $R$ is a measure of the relative
anisotropy variations throughout the spectrum and should at least
qualitatively characterize the spectral behavior of this anisotropy. It is
therefore meaningful to compare the $R$ spectrum with the $Q/I$
spectrum (the Second Solar Spectrum) to look for qualitative
similarities. It turns out that the two spectra have little in
common. 

%%%%%%%%%%%%%%%%%%%%%%%%%
\subsection{Terminology: SS3}\label{sec:SS3}
Before doing this comparison, let us introduce a terminology that is
the most convenient to use 
when referring back and forth between the various
types of spectra. When the spectral richness of the linearly polarized
spectrum that is produced by coherent scattering processes became
apparent, the term Second Solar Spectrum was introduced \citep{stenflo-ivanov91,stenflo-sk97}, because the new spectrum had little resemblance
with the intensity spectrum and was largely governed by other physical
processes. It was like being confronted with an entirely new and
unfamiliar spectral face of the Sun. This choice of name (which
became the widely adopted terminology soon after it was introduced)
indirectly implies that 
we think of the ordinary intensity spectrum as the First Solar
Spectrum. Similarly our ratio spectrum
$R_\lambda$ differs profoundly from both the intensity spectrum and
the Second Solar Spectrum while being as rich in
structures. $R$ cannot be derived from either $I$ or $Q/I$, it is
not governed by the same physics. We therefore find it natural to
think of it as the Third Solar Spectrum, another new spectral face of the
Sun. 

When frequently referring to these various spectra it is however more
convenient to make use of acronyms. In the following we
will refer to the intensity spectrum as SS1, the Second Solar Spectrum
as SS2, and our ratio spectrum $R$ as SS3. All three of them are
functions of $\mu$. 

\begin{figure}
\resizebox{\hsize}{!}{\includegraphics{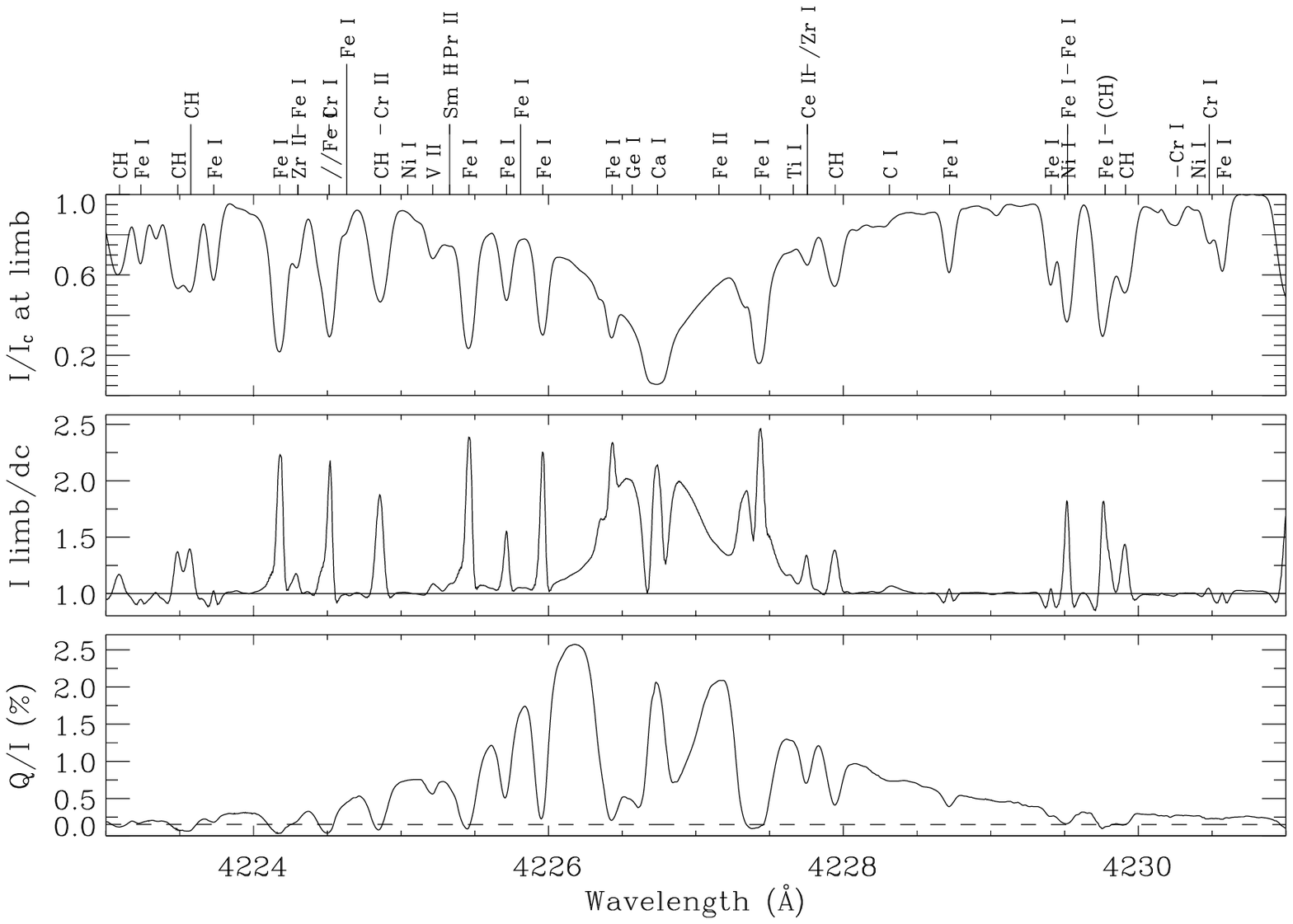}}
\caption{Comparison between three profoundly different spectral
  faces of the Sun, for the spectral window 4223-4231\,\AA\ around
  the strong Ca\,{\sc i} 4227\,\AA\ line. Top panel: SS1 (the
  intensity spectrum) for $\mu=1.0$. Middle panel: SS3 (the
  limb/disk-center ratio spectrum $R$) for $\mu=0.145$. Bottom panel:
  SS2 (the Second Solar Spectrum) for $\mu=0.1$ with the level of the
  continuum polarization drawn as the horizontal dashed line. 
}\label{fig:4227}
\end{figure}

%%%%%%%%%%%%%%%%%%%%%%%%%
\subsection{Comparison between SS3 and SS2}\label{sec:comparess}
Let us next compare the appearance of the three spectra with each
other for a few selected spectral windows. The data used for SS2 are
based on Volumes I and II of the Atlas of the Second Solar Spectrum
\citep{stenflo-gandorf00,stenflo-gandorf02}, here converted into digital form, shifted to
conform to the continuum polarization and zero point of the
polarization scale as determined in \citet{stenflo-s05}, and with a
mild, conservative application of wavelet smoothing. The whole SS2
Atlas, from 3161 to 6987\,\AA, is available in this form both as pdf
and data files at the IRSOL web site www.irsol.ch. 

Figure \ref{fig:4227} shows a section around the Ca\,{\sc i}
4227\,\AA\ line, which has the largest scattering polarization
amplitude in the whole visible solar spectrum. SS1 at disk center is
shown in the top panel, SS3 for $\mu=0.145$ in the middle panel, SS2 for
$\mu=0.1$ in the bottom panel. SS2 and SS3 differ greatly in both the position
of the maxima and minima within the Ca profile, and in the profile
width. In SS2 the blend lines depolarize the Ca polarization down to
the continuum polarization level, while in SS3 the blend lines appear as peaks. 

According to Eq.~(\ref{eq:pfact}), SS2 that is represented by $p$
should be compared with $k_G$ rather than with SS3 
(that is represented by our ratio spectrum $R$). Since we only have $R$ for a
single $\mu$ position we cannot properly calculate $k_G$ here, but
we know that $k_G$ is similar to an inverted version of $R$, because when $R$ goes
up like an emission line, $k_G$ goes down. This can be understood from
Eq.~(\ref{eq:crc}). For the wavelength 4227\,\AA, ${\cal C}_c =0.29$,
representing a reduction of the continuum intensity relative to disk
center by a factor of 3.44. This factor is largely compensated for
by the SS3 factor $R$, which according to the middle panel of
Fig.~\ref{fig:4227} is approximately 2, which elevates the net
limb-darkening function ${\cal C}$ within the line from 0.29 to about
0.6 at this limb position. The limb darkening function is thus much
flatter inside the lines, implying a {\it smaller} anisotropy $k_G$. 

As $k_G$ is largest in the continuum, {\it all} the lines will
appear like absorption lines in the $k_G$ spectrum. In contrast, all
the polarizing lines appear like emission lines in SS2. If we would
replace the SS3 spectrum with the $k_G$ spectrum, the resemblance with
the SS2 spectrum would rather get worse than better. 

For the particular spectral window around the 4227\,\AA\ line a
high-resolution $k_G$ spectrum has been published in
\citet{stenflo-sampoorna09}, based on recordings with ZIMPOL at IRSOL
for a sequence of $\mu$ positions on the quiet Sun. Inspection of this
plot verifies that it is indeed similar to an absorption-like inverted version of
SS3 in Fig.~\ref{fig:4227}. 

\begin{figure}
\resizebox{\hsize}{!}{\includegraphics{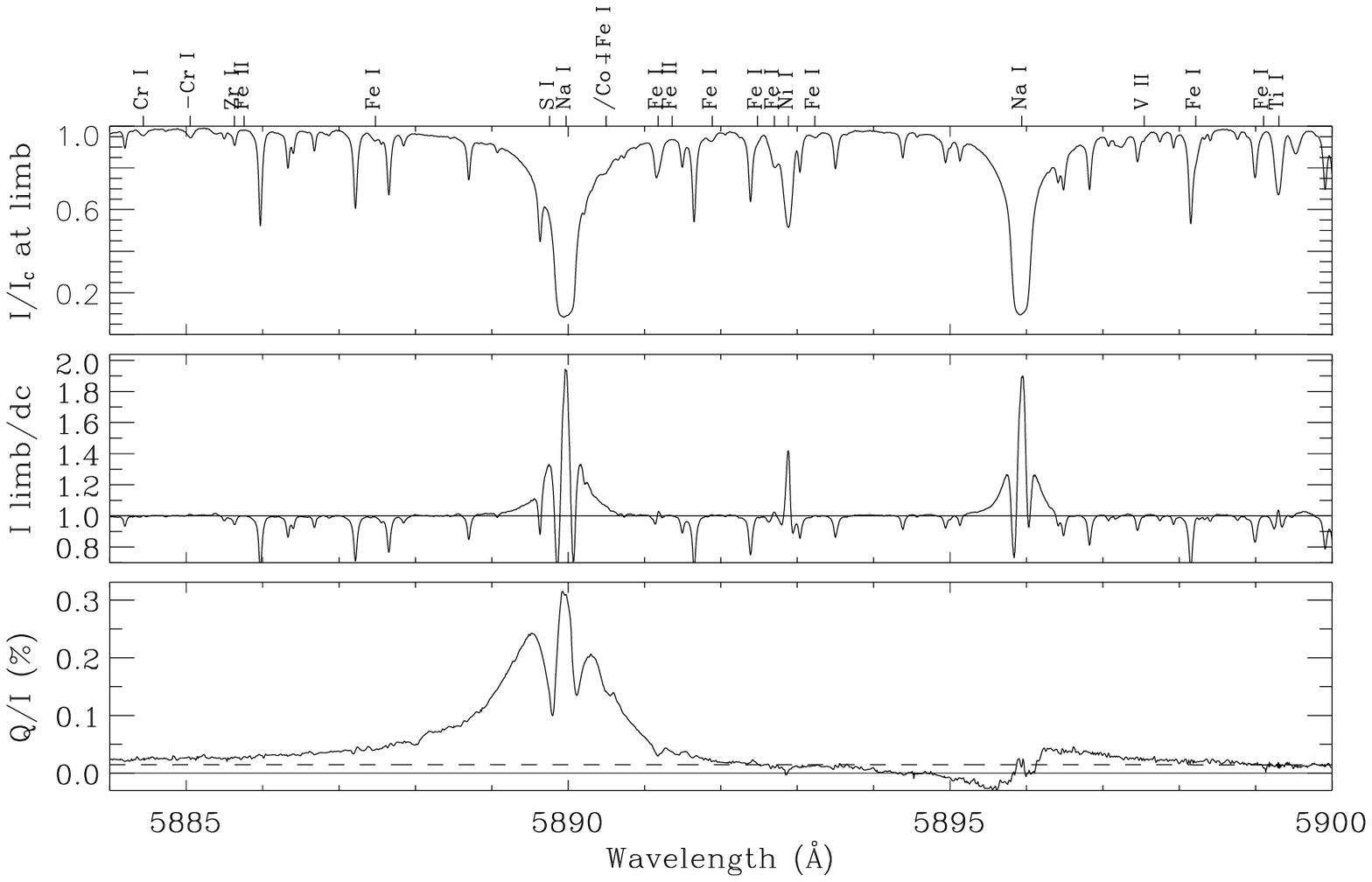}}
\caption{Same as Fig.~\ref{fig:4227}, but for the spectral range 5884
  - 5900\,\AA\ around the well known Na\,{\sc i} D$_2$ and D$_1$ lines
  at 5890 and 5896\,\AA. 
}\label{fig:NaD2D1}
\end{figure}

Figure \ref{fig:NaD2D1} shows as a second example the region around
the famous Na\,{\sc i} D$_2$ and D$_1$ lines at 5890 and
5896\,\AA. While the D$_2$ line might appear  to have a qualitatively
similar line shape in SS2 and SS3, the line widths and positions of
the side peaks are very different. With the D$_1$ line there is not a
trace of similarity.  

\begin{figure}
\resizebox{\hsize}{!}{\includegraphics{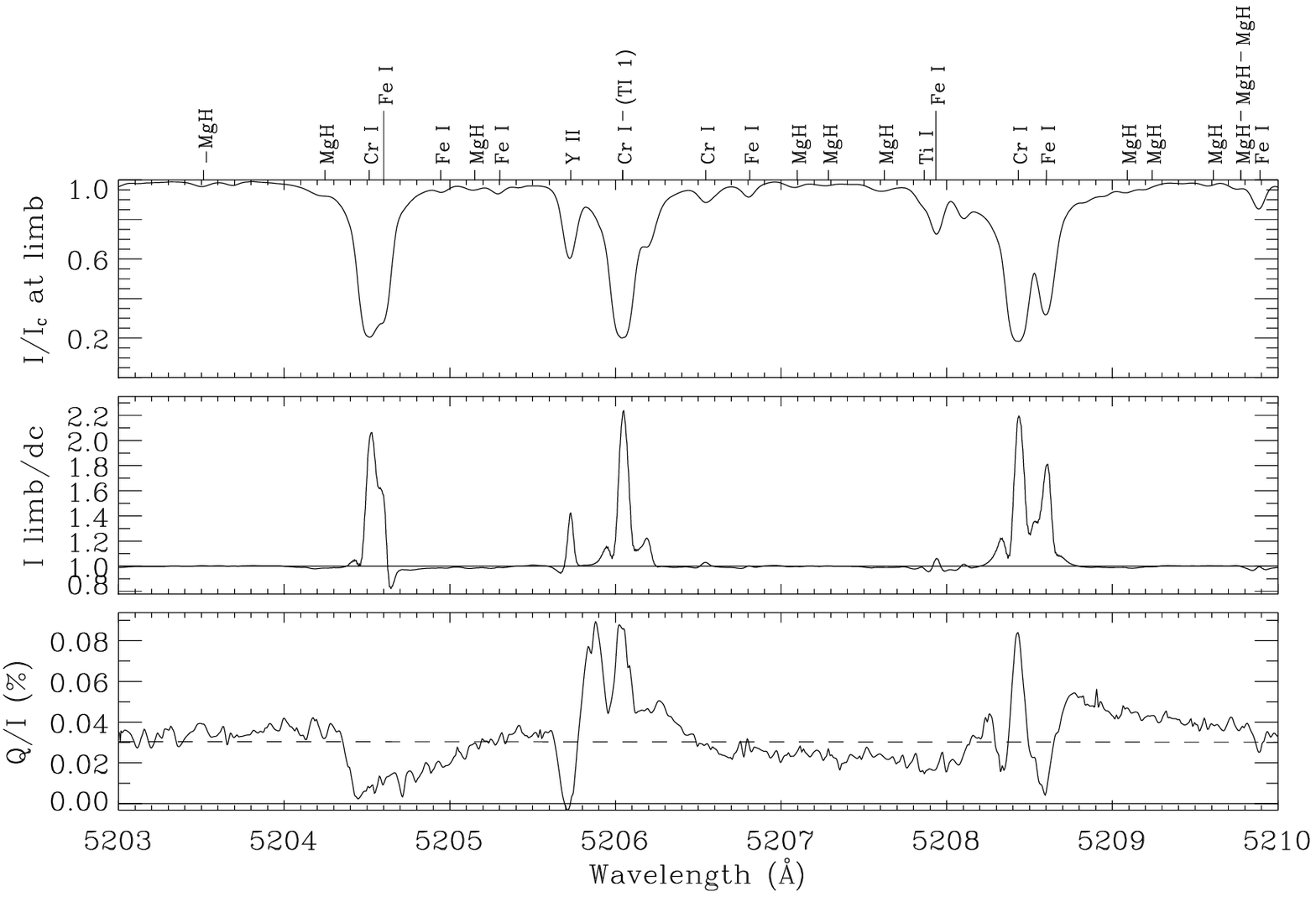}}
\caption{Same as Fig.~\ref{fig:4227}, but for the spectral range 5203
  - 5210\,\AA\ that contains the Cr\,{\sc i} triplet lines, which in
  SS2 exhibit dramatic signatures of quantum interferences between the
  atomic states of different total angular momenta. 
}\label{fig:Crtriplet}
\end{figure}

The complete absence of any similarity is more strikingly demonstrated
in Fig.~\ref{fig:Crtriplet} for the Cr\,{\sc i} triplet lines in the
range 5203 - 5210\,\AA. These three chromium lines represent fine
structure components with different $J$ quantum numbers within the
same multiplet, with quantum interferences between the different $J$
states which dramatically reveal themselves in the form of the
remarkable polarization pattern seen in SS2. This interference pattern has been
modeled in great detail with polarized radiative transfer and partial
frequency redistribution by \citet{stenflo-smitha12}. The three panels
in Fig.~\ref{fig:Crtriplet} demonstrate that we are dealing with three
distinctly different spectra, which depend on the physics of the Sun
in different ways. 

Since the anisotropy appears as a scaling factor for the scattering
polarization in Eq.~(\ref{eq:pfact}), it may seem surprising that
there is so little resemblance between SS2 and SS3. There are several
reasons for this.  One is that the intrinsic polarizability
$W_{2,\,{\rm eff}}$ is governed by quantum physics that has nothing to
do with SS3. Another is that the SS2 profiles are largely shaped by
polarized radiative transfer with partial frequency redistribution,
and that the contributing incident radiation field has an anisotropy
that varies along the line of sight. The most probable optical depth
(along the line of sight) from which the observed photons originate is
$\tau=1$. In contrast, SS3 reflects the properties of the anisotropy
at the surface, at $\tau=0$. 

Let us further recall that Eq.~(\ref{eq:pfact}) represents the
idealized case of single scattering by a particle sitting on top of
the atmosphere, being illuminated from below. This idealization is
often referred to as the Last Scattering Approximation (LSA) and
is useful for conceptual discussions and for crude estimates of the
expected polarization amplitudes without the need to go into the
complexities of polarized radiative transfer. LSA has played an important role in
the discovery and modeling of SS2 signatures of quantum interferences
between states of different total angular momenta \citep{stenflo-s80} and
for initial estimates via the Hanle effect that the strength of the
hidden, microturbulent magnetic fields that fill the photosphere 
lies in the range 10-100\,G \citep{stenflo-s82}. The LSA
concept has also been applied and greatly generalized for the
interpretation of SS2 line profiles by \citet{stenflo-anusha10}. The lack of
resemblance between SS2 and SS3 however demonstrates the limitations
of the LSA idealization and indicates that much more complete radiative-transfer
modeling is required to quantitatively understand and interpret the
complex profile structures in SS2.

%%%%%%%%%%%%%%%%%%%%%%%%%%%%%%%%%%%%%%%%%%%%%%%%%%%%%%%%%%%%%
\section{Model of the CLV in terms of the intensity
  spectrum at disk center}\label{sec:model}
In Sect.~\ref{sec:clvatlas} we noticed that there might exist a
relatively well defined non-linear relation between SS1 and SS3 in
the case of the weak to medium-strong lines, which are largely
governed by LTE processes that are directly coupled to the local
temperature-density structure of the atmosphere. In the present
section we will try to model 
the relation between SS1 and SS3 for the LTE lines. The model 
  does not apply to the stronger (or to many of the intermediately
  strong) lines, since we know that such lines can only be
  successfully reproduced in terms of 3-D models and non-LTE physics. 

Comparison between the SS1 and SS3 profile shapes in
Fig.~\ref{fig:4391} indicates that the following model might work, a
conjecture that gets vindicated when it is subsequently applied to our
atlas data set. Let us for convenience introduce the relative line
depth $d_\lambda$ for the disk-center spectrum (SS1): 
\begin{equation}\label{eq:d}
d_\lambda=1-r_\lambda(\mu=1.0)\,,
\end{equation}
where $r_\lambda$ is the relative rest intensity (in units of the
continuum intensity) as in Eq.~(\ref{eq:restint}). The proposed model $R_{\rm
  model}$ that represents how SS1 gets converted into SS3, the ratio
spectrum $R$, can then be expressed as 
\begin{equation}\label{eq:rmodel}
R_{\lambda,\,{\rm  model}}= A\,d_\lambda^\beta\,+1\,.
\end{equation}
The number 1 is added, because in the absence of lines
($d_\lambda=0$) the continuum level $R=1$ must be retrieved. 

Note that in Eq.~(\ref{eq:rmodel}) $d_\lambda$ refers to disk
  center and not to the $\mu$ position of the $R$ spectrum. This
  implies that the CLV behavior, represented by $R$ on the left-hand
  side, is modeled exclusively in terms of the disk-center spectrum on
the right-hand side. Conceptually this means that the whole CLV
behavior of the LTE-type lines can be directly inferred, once we know the
disk-center spectrum. 

\begin{figure}
\resizebox{\hsize}{!}{\includegraphics{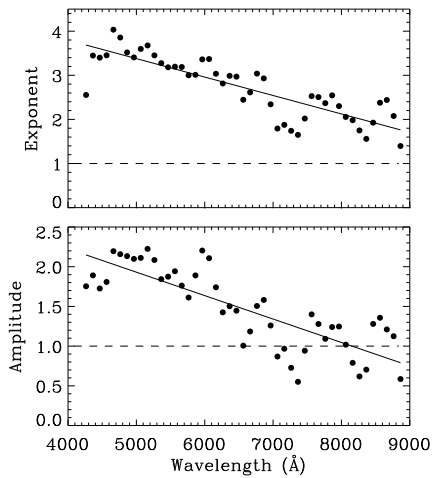}}
\caption{Parameters $\beta$ (exponent, top panel) and $A$ (amplitude,
  bottom panel) as obtained from  least squares fitting of SS3 with the model of 
  Eq.~(\ref{eq:rmodel}). The straight-line fits to the values of
  $\beta$ and $A$ are given by Eqs.~(\ref{eq:beta}) and
  (\ref{eq:apar}). $\beta=1$ would represent a linear relation between
  SS1 and SS3. $\beta -1$ therefore expresses the degree of non-linearity.
}\label{fig:modelfit}
\end{figure}

This model is then applied with an iterative least squares procedure
to determine the two free parameters $A$ and $\beta$ for a sequence of
10\,\AA\ wide spectral windows, incrementally shifted by 5\,\AA\
relative to each other, but
after excluding the sections around the strong and wide lines that we
know do not conform to our model. As the fits to the various spectral
sections are of varying quality because all admitted lines do not fit
the model equally well, we retain only the well-behaved fits by
rejecting those for which the standard error in $A$ and $\beta$ falls outside
the range 0.4-1.6 times the median value of the standard error. Trial
and error shows that when
doing so the scatter (between the results for the different spectral
sections) of the fit parameters $A$ and $\beta$ decreases, which is
the main reason why we call these fits more well-behaved. To further
reduce the scatter we spectrally smooth $A$ and $\beta$ with a running
200\,\AA\ wide window and sample the smoothed function in steps of
100\,\AA. The result is illustrated in Fig.~\ref{fig:modelfit}. 

Next we have fitted straight lines through the points in the two
panels of Fig.~\ref{fig:modelfit}. We find that the straight-line fits
do not significantly depend on whether the fits to the $\beta$ and $A$
values are done before or after the mentioned 200\,\AA\ smoothing. Also tests with
second-order polynomial fits give no significant curvature
(second-order) term, there is no reason to go beyond first order with
the present data set. The two lines are given by 
\begin{equation}\label{eq:beta}
\beta= 5.47\,  -0.42\,   (\lambda/1000)
\end{equation}
 and
\begin{equation}\label{eq:apar}
A= 3.41\, -0.30\, (\lambda/1000)\,,
\end{equation}
where $\lambda$ should be given in units of \AA. 

Of particular significance is the magnitude of $\beta$, since $\beta-1$
expresses the degree of non-linearity in the relation between SS1 and
SS3. The relation becomes increasingly non-linear as we go to shorter
wavelengths. This behavior most probably reflects a fundamental
property of the solar atmosphere, which valid model atmospheres need to
be able to reproduce. 

\begin{figure}
\resizebox{\hsize}{!}{\includegraphics{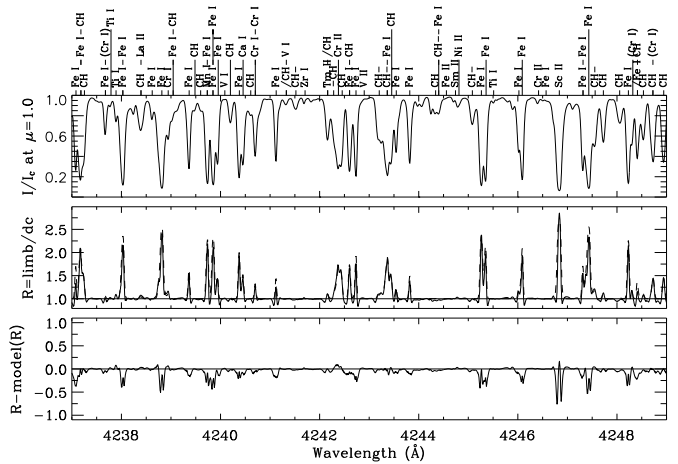}}
\caption{Comparison, for the 12\,\AA\ wide spectral range 4237-4249\,\AA, between
  the observed limb/disk-center ratio spectrum $R$ (SS3, solid curve
  in the middle panel) and the model of 
  $R$, overplotted as the dashed curve. The bottom panel gives the
  difference between the solid and dashed curves, on the same scale as
  used in the middle panel, to allow a direct comparison. The top
  panel shows the intensity spectrum (SS1) at disk center with line
  identifications. 
}\label{fig:4245model}
\end{figure}

\begin{figure}
\resizebox{\hsize}{!}{\includegraphics{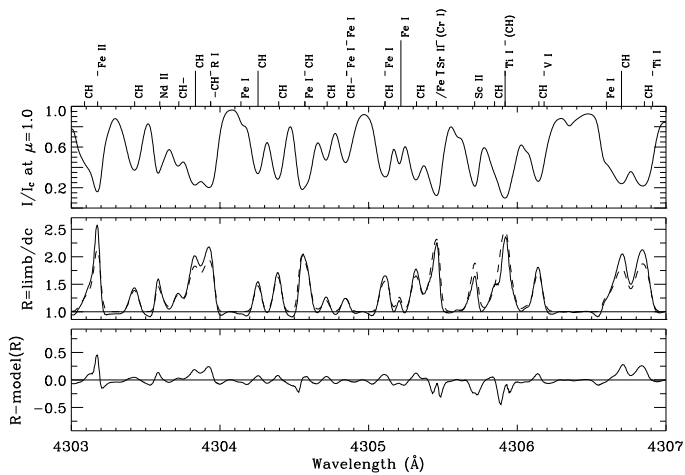}}
\caption{Same as Fig.~\ref{fig:4245model} but for the 4\,\AA\ wide
  spectral range 4303-4307\,\AA, to allow a more detailed comparison
  between the actual and the modeled profile shapes. 
}\label{fig:4305model}
\end{figure}

Figures \ref{fig:4245model} and \ref{fig:4305model} illustrate two
examples of spectral sections where the model works particularly
well. In other parts of the spectrum there are many lines that do not
conform that well to the model, but for most weak to medium-strong lines
the given examples provide a rather representative picture of the
goodness of the fit. As before, the top panel shows the disk-center
spectrum with line identifications. The observed ratio spectrum $R_\lambda$ (SS3) is
given as the solid curve in the middle panel, with the model value
$R_{\lambda,\,{\rm  model}}$ from Eqs.~(\ref{eq:rmodel})-(\ref{eq:apar}) overplotted as 
the dashed curve. Since the solid and dashed curves agree so well that
it is hard to distinguish them from each other, we plot in the bottom
panel the difference $R_\lambda\,-R_{\lambda,\,{\rm  model}}$ between
the observed and modeled ratio spectrum, using the same plot scale as in
the middle panel to allow a direct comparison of the model
deviations with the amplitudes of the original SS3 spectrum. 

The remarkable success of our very simple analytical model that describes the
translation from SS1 to SS3 for weak to medium-strong, LTE-type lines
is a consequence of some fundamental property of the temperature-density
structure of the solar atmosphere, which governs the relation between
the intensity spectrum and its center-to-limb variation. The values
and wavelength variations of the model parameters $\beta$ and $A$, as
given by Eqs.~(\ref{eq:beta}) and  (\ref{eq:apar}), therefore
constitute in a highly compact form a set of novel observational
constraints on model atmospheres. 

Although the discovery that the relation between SS1 and SS3 may be
modeled with such a simple analytical expression as that of
Eq.~(\ref{eq:rmodel}) was unexpected, it is not too surprising that
the two spectra are physically related in a rather well-defined way
for the parts of the spectrum that can be described in terms of LTE
physics. In this case both the intensity spectrum and its center-to-limb
variation directly depend on the local temperature-density
structure of the atmosphere. They have a common origin, although the
dependence on the atmospheric structures plays out in different ways
for SS1 and SS3. The situation is different for SS2, which is governed
by non-LTE physics and quantum phenomena which have no well-defined
counterparts in 
SS1 or SS3. Therefore it is not possible to relate SS2 to SS1 or SS3 
by phenomenological models, it is a profoundly different kind of
spectrum.

%%%%%%%%%%%%%%%%%%%%%%%%%%%%%%%%%%%%%%%%%%%%%%%%%%%%%%%%%%%%%
\section{Conclusions}\label{sec:conclude}
The temperature-density stratification of the Sun's atmosphere reveals
itself in the way in which the intensity varies across the solar
disk. The variation with heliocentric angle $\theta$ can also be
interpreted as the angular distribution of the radiation field at the
surface, where $\theta$ is the inclination angle of the radiation with
the vertical direction. The intensity spectrum varies with disk
position as labeled by $\mu=\cos\theta$. The spectral structuring of
this variation is best brought out by normalizing the spectrum for a
given $\mu$ with its counterpart at disk center, i.e., forming the
ratio $R_\lambda(\mu)=r_\lambda(\mu)/r_\lambda(1.0)$ between the
respective continuum-normalized spectra
$r_\lambda=I_\lambda/I_{c,\,\lambda}$. 

In the present paper we have presented an FTS atlas of $R_\lambda$ at
$\mu=0.145$, representing a disk position on the quiet Sun 10\,arcsec
inside the limb. The wavelength range 4084 - 9950\,\AA\ is covered
with high S/N ratio, full spectral resolution, and no spectral stray
light. The entire atlas is available as both pdf and data files at
the web site www.irsol.ch of IRSOL, Locarno. In a separate study,
carried out at IRSOL with ZIMPOL, a sequence of spectral atlases
of $R_\lambda$ for the quiet Sun  are being produced. They will
represent 10 disk positions that are equidistant in $\mu$, from disk center to 
5\,arcsec inside the limb, and will thereby complement
and extend the present work. However, although the present atlas of
$R_\lambda$ only represents the single disk position 
$\mu=0.145$, it reveals the basic rich spectral structuring of the Sun's
center-to-limb variation. 

As we are here dealing with several qualitatively different
representations of the Sun's spectrum, it is convenient to refer to
them with the acronyms SS1, SS2, and SS3, thereby extending the previous
terminology Second Solar Spectrum (here represented by the acronym
SS2) to the qualitatively new $R_\lambda$ spectrum, which we refer to
as SS3 (more explicitly as the Third Solar Spectrum). All three
spectra depend on $\mu$, but the amplitudes of the spectral structures
in both SS2 and SS3 increase monotonically as we move towards the
limb. 

Since SS3 can be seen as a spectral representation of the anisotropy
of the emergent radiation field, and since this anisotropy is a source
of the scattering polarization that is mapped by SS2, one might expect
that there may be some similarities between SS2 and SS3. A 
comparison between them shows however that they are structured in totally
different ways. SS2 is governed by non-LTE physics and quantum processes
that do not leave discernible signatures in SS1 or SS3. In contrast we
discover the existence of a mapping relation between SS1 and SS3
in the case of the weak and medium-strong lines that are primarily
governed by LTE physics. Although this mapping is highly non-linear,
we succeed to model it in terms of a simple power law
governed by two free parameters (amplitude and exponent), which vary
approximately linearly across the entire wavelength range covered by
our atlas. The values and wavelength variations of these two free
parameters represent a new set of observational constraints on models
of the quiet Sun. 

Next we plan to apply the SS3 constraints to test the
validity of existing model atmospheres and explore how they
should be improved. Such modeling work will also advance the
quantitative interpretations of the SS2 line 
profiles and their center-to-limb variations. Recent SS2
modeling with elaborate polarized radiative-transfer and the use
of grids of realistic 1-D model atmospheres has failed to reproduce
key features of selected SS2 line profiles, which has led to the
suggestion that it may be necessary to go beyond 1-D models to
multi-dimensional geometries \citep{stenflo-supriya14}. The 
multi-dimensional models  can be tested through comparison with 
our SS3 constraints after the computed emergent radiation has been
spatially averaged in the horizontal plane. Such tests are much easier
to do than SS2 modeling, since they are done for unpolarized radiation. 
In general the 3-D hydrodynamic and MHD models may be refined
  this way. The spectral CLV data may also help calibrating the 
  badly known cross sections of the collisions with neutral hydrogen
  atoms that are used in non-LTE calculations.

\bibliographystyle{aa}
%\bibliography{../stenflo_aa-2014}

\end{document}